\documentclass[a4paper]{jpconf}
\usepackage{graphicx}
\usepackage{iopams}

\newcommand {\mofe} {\{$\textrm{Mo}_{72}\textrm{Fe}_{30}$\}}
\newcommand{\op}[1]{%
    \fontdimen12\textfont3=2pt\fontdimen12\scriptfont3=1.4pt%
    \!\null\mathop{\vphantom{#1}\smash{#1}}\limits_{\sim}\null\!}

\def\ket#1{\, | \, {#1} \, \rangle}

\newcommand{\figref}[1]{Fig.~\protect\ref{#1}}
\newcommand{\pp}[2]{\frac{\partial \, {#1}}{\partial \, {#2}}}
\newcommand {\Emin} {E_{\rm min}}
\newcommand {\Bsat} {B_{\rm sat}}

\begin{document}
\title{Exact eigenstates of highly frustrated spin lattices probed in high fields}

\author{J Schnack$^1$, H-J Schmidt$^1$, A Honecker$^2$, J Schulenburg$^3$ and J Richter$^4$}

\address{$^1$ Dept. of Physics, University of Osnabr\"uck, Barbarastr. 7, 49069 Osnabr\"uck, Germany}
\address{$^2$ Inst. of Theor. Phys., University of G\"ottingen, Fr.-Hund-Platz 1, 37077 G\"ottingen, Germany}
\address{$^3$ URZ, University of Magdeburg, 39016 Magdeburg, Germany}
\address{$^4$ FNW/ITP, University of Magdeburg, PF 4120, 39016 Magdeburg, Germany}
\ead{jschnack@uos.de}

\begin{abstract}
  Strongly frustrated antiferromagnets such as the magnetic
  molecule \mofe, the kagome, or the pyrochlore lattice exhibit
  a variety of fascinating properties like low-lying singlets,
  magnetization plateaus as well as magnetization jumps. During
  recent years exact many-body eigenstates could be constructed
  for several of these spin systems. These states become ground
  states in high magnetic fields, and they also lead to exotic
  behavior. A key concept to an understanding of these
  properties is provided by independent localized magnons. The
  energy eigenvalue of these $n$-magnon states scales linearly
  with the number $n$ of independent magnons and thus with the
  total magnetic quantum number $M=Ns-n$. In an applied field
  this results in a giant magnetization jump which constitutes a
  new macroscopic quantum effect. It will be demonstrated that
  this behavior is accompanied by a massive degeneracy, an
  extensive $(T=0)$-entropy, and thus a large magnetocaloric
  effect at the saturation field. The connection to flat band
  ferromagnetism will be outlined.
\end{abstract}

\section{Introduction}
\label{sec-1}
Geometric frustration of interacting spin systems is the driving
force of a variety of fascinating phenomena in low-dimensional
magnetism \cite{FRUSTRATION}. 
In this context the term \emph{frustration} describes a situation
where in the ground state of a classical spin system not all
interactions can be saturated simultaneously. A typical picture
for such a situation is a triangle of antiferromagnetically
coupled spins, where classically the spins are not in the
typical up-down-up configuration, but assume a ground state that
is characterized by a relative angle of $120^\circ$ between
neighboring spins.  This special classical ground state
characterizes among others several frustrated spin systems which
are built of corner-sharing triangles, among them the giant
Keplerate molecule \mofe\ which is a perfect icosidodecahedron
\cite{MLS:CPC01} and the kagome lattice antiferromagnet.  The
pyrochlore lattice, which consists of corner-sharing tetrahedra
and thus has a different structure, nevertheless shares several
important properties with the above systems.

Research in this field is naturally focused on the low-energy,
i.e.\ low-temperature, low-field behavior. A key observation is
that the quantum spin systems possess many or even infinitely
many singlet states below the first triplet state and the
classical counterpart systems display a non-trivial ground state
degeneracy \cite{MoC:PRB98}. Another important observation
concerns a plateau in the magnetization curve for $T=0$, which
for the systems made of corner-sharing triangles is at
${\mathcal M} = {\mathcal M}_{\rm sat}/3$, see e.g.
\cite{KaM:JPSJ85,Zhi:PRL02}, whereas for the pyrochlore it is at
${\mathcal M} = {\mathcal M}_{\rm sat}/2$, see e.g.
\cite{PSS:PRL04}.

In this article we focus on special properties of these systems
which arise at low temperatures but high magnetic fields. We
will show that it is possible to construct exact many body
states which are product states of independent one-magnon states.
These states become ground states in high magnetic fields.  In a
wider perspective such an arrangement of independent
single-particle objects can be understood as condensation of
bosons \cite{SSR:EPJB01,ZhT:PRB04,ZhT:IKYS05}. The linear
scaling of the minimal total energy with the number of these
objects explains their unusual high-field behavior which is
expressed in magnetization jumps \cite{SHS:PRL02,RSH:JPCM03},
non-zero $(T=0)$-entropy \cite{DeR:PRB04,HoR:CMP05}, and an
enhanced magnetocaloric effect
\cite{ZhH:JSM04,DeR:06}.

\section{Concept of independent magnons}
\label{sec-2}
In the following we assume that the spin systems under
consideration are modeled by an isotropic Heisenberg Hamiltonian
augmented with a Zeeman term, i.e.,
\begin{eqnarray}
\label{E-2-1}
\op{H}
&=&
-
\sum_{u, v}\;
J_{uv}\,
\op{\vec{s}}(u) \cdot \op{\vec{s}}(v)
+
g \mu_B B \op{S}_z
\ .
\end{eqnarray}
$\op{\vec{s}}(u)$ are the individual spin operators at sites
$u$ and $\op{S}_z$ is the $z$-component of the total spin.
$J_{uv}$ are the matrix elements of the symmetric coupling
matrix. We will consider only antiferromagnetic couplings.

\begin{figure}[h]
\begin{minipage}{18pc}
\begin{center}
\includegraphics[width=14pc]{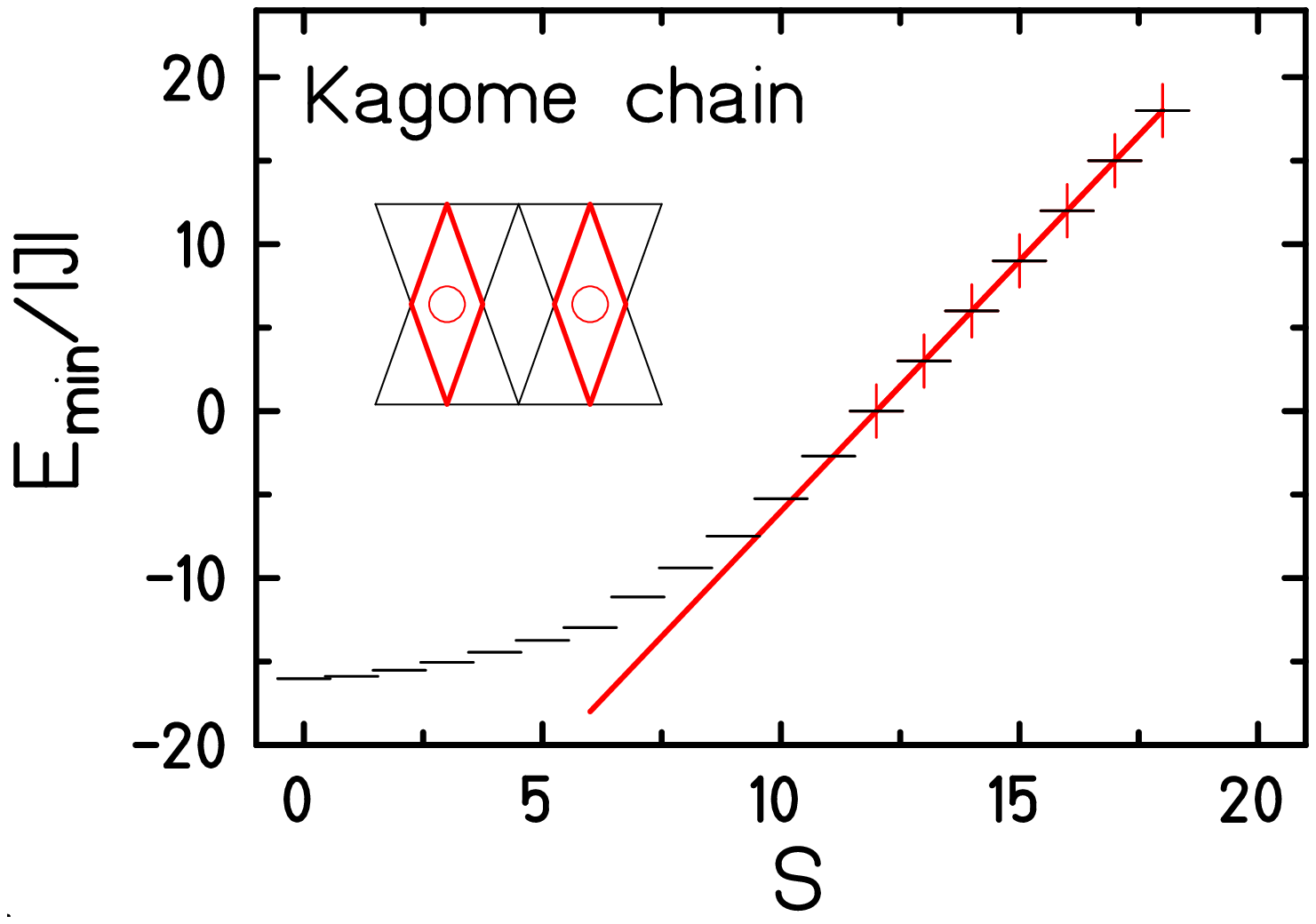}
\caption{\label{F-1-A}Minimal energies $\Emin$ of a kagome chain
  with $N=36$ and $s=1/2$. The highest seven levels fall on a
  straight line. The highlighted diamonds in the structure are
  localized magnons.}
\end{center}
\end{minipage}\hspace{2pc}%
\begin{minipage}{18pc}
\begin{center}
\includegraphics[width=14pc]{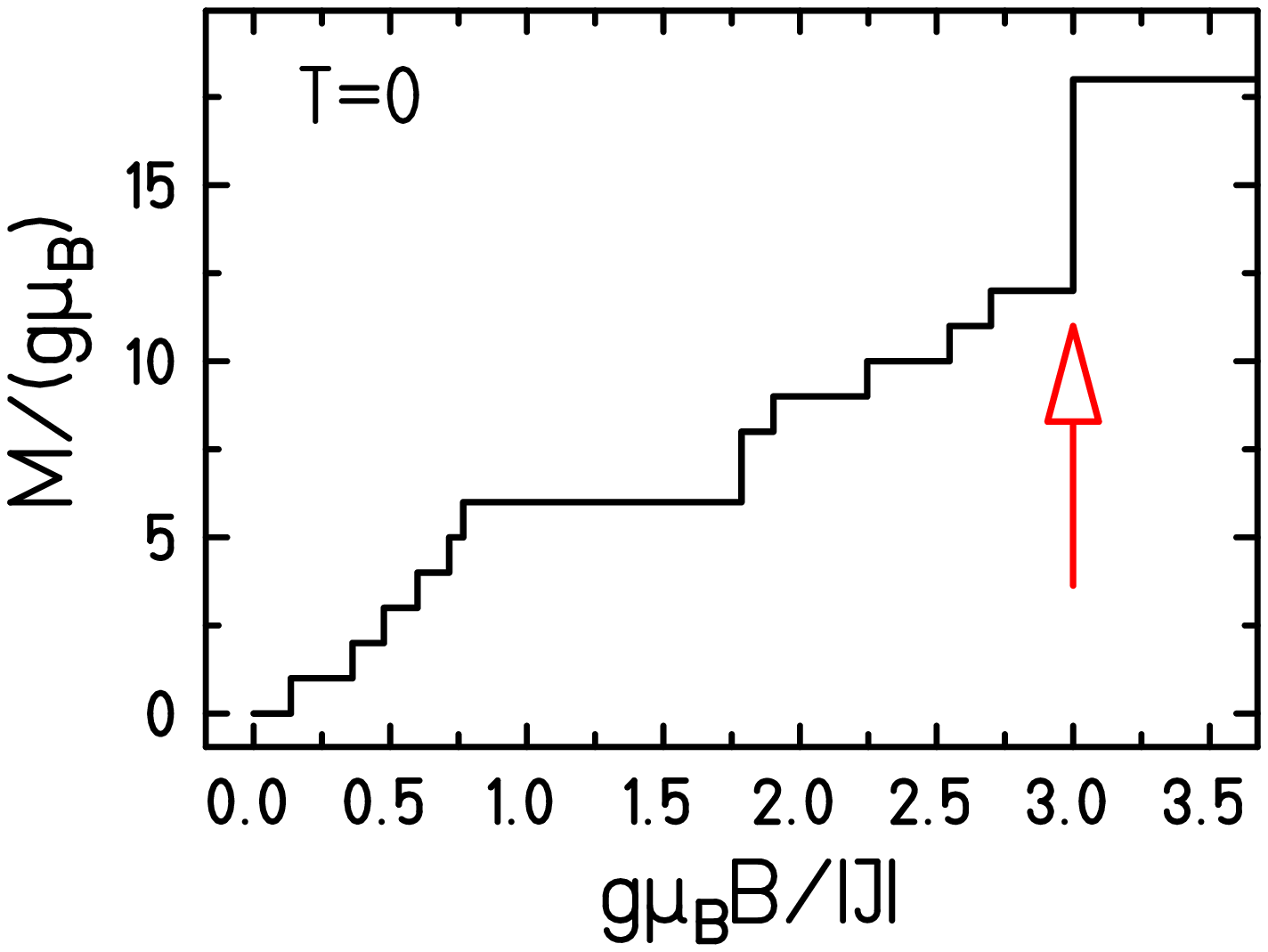}
\caption{\label{F-1-B}$(T=0)$-magnetization curve of the kagome
  chain, \figref{F-1-A} with $N=36$.
  The magnetization jump of $\Delta M=6$ is marked by an arrow.}
\end{center}
\end{minipage} 
\end{figure}

One of the early unexpected results was that for the
icosidodecahedron (e.g. \mofe) the minimal energies $\Emin$ in
each Hilbert subspace of ${\mathcal H}(M)$ of total magnetic
quantum number $M$ scale linearly with $M$ close to the
saturation \cite{SSR:EPJB01}.  Figure \ref{F-1-A} shows as
another example the minimal energy levels of a special kagome
chain (introduced in \cite{Wal:PRB00})
where the seven highest levels are on a straight line. In
an applied magnetic field this leads to a simultaneous crossing
of the lowest Zeeman levels at the saturation field, which gives
rise to an unusual magnetization jump, that for the kagome
chain is shown in \figref{F-1-B}.

\begin{figure}[h]
\begin{minipage}{18pc}
\begin{center}
\includegraphics[width=8pc]{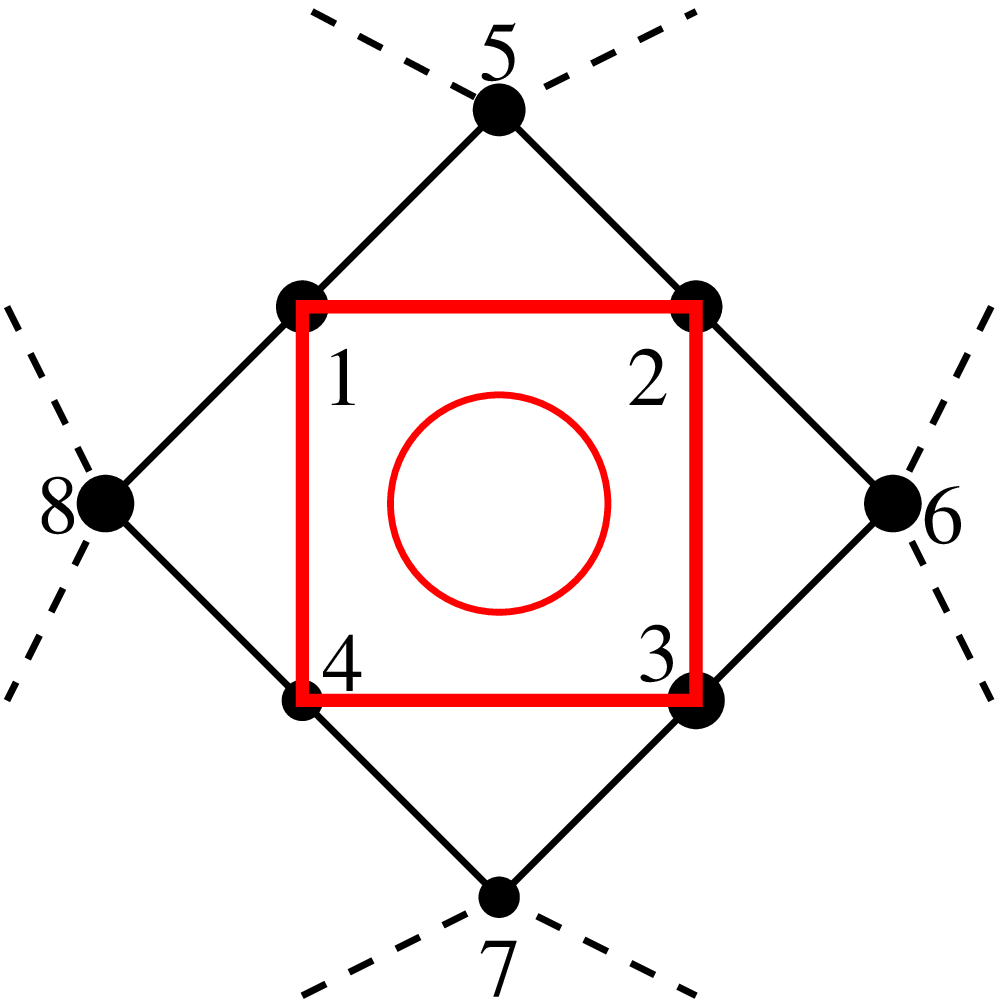}
\caption{\label{F-2-A}Localized one-magnon state on a model
  lattice.}
\end{center}
\end{minipage}\hspace{2pc}%
\begin{minipage}{18pc}
\begin{center}
\includegraphics[width=8pc]{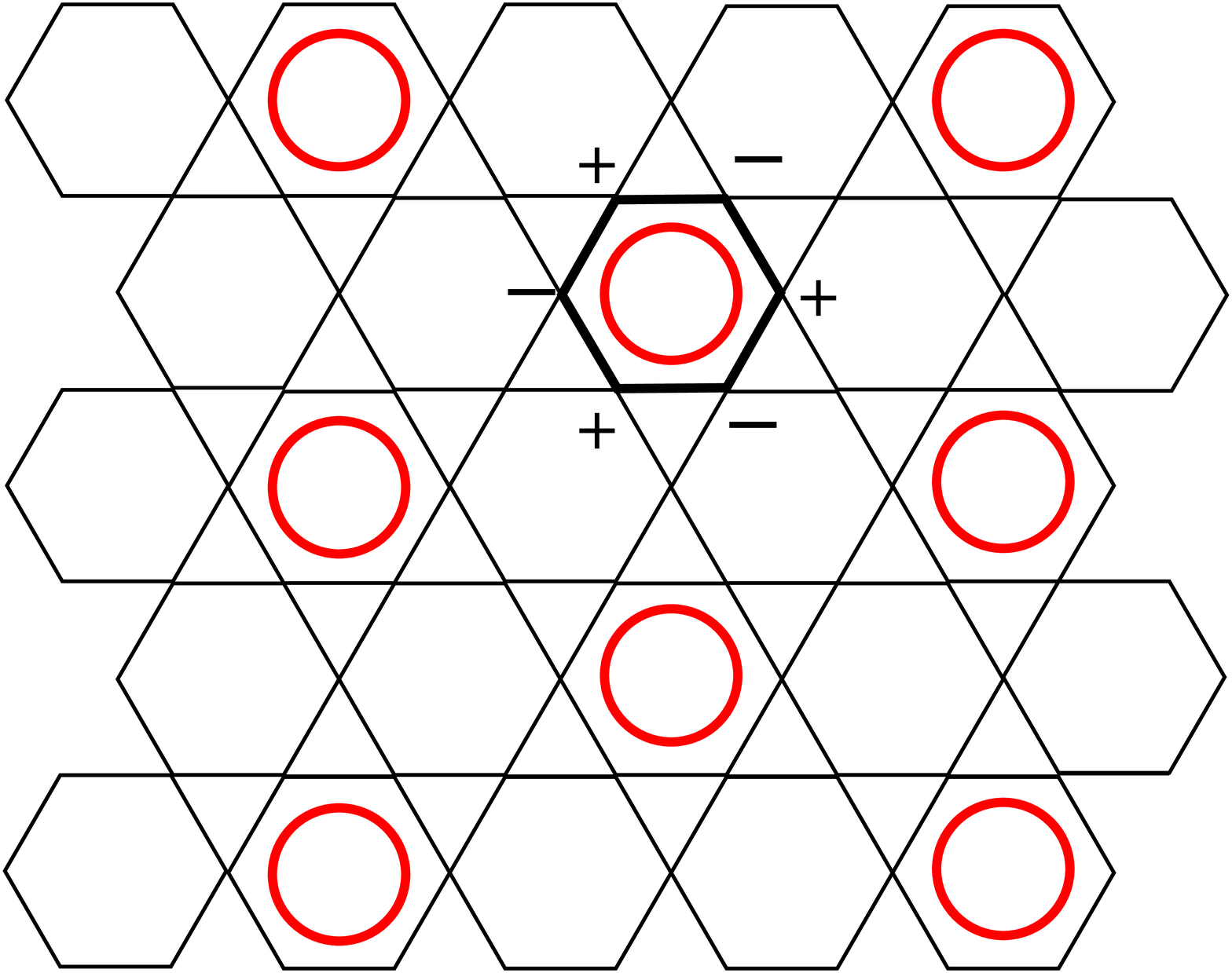}
\caption{\label{F-2-B}Independent localized magnons on the
  kagome lattice antiferromagnet. $N/9$ independent localized
  magnons can be placed on the kagome lattice.}
\end{center}
\end{minipage} 
\end{figure}

It turns out that the linear dependence of $\Emin$ on $M$ can be
understood in terms of localized independent magnons. Figure
\ref{F-2-A} shows the structure of a localized magnon on a part
of a model lattice as realized for instance in the kagome chain
of \figref{F-1-A}. The one-magnon state is given by
\begin{eqnarray}
\label{E-2-2}
\ket{\mbox{localized magnon}}
&=&\frac{1}{2}\left(\ket{1}-\ket{2}+\ket{3}-\ket{4}\right)
\ ,\quad\mbox{where}
\\
\ket{1}
&=&
\op{s}^{-}(1)\ket{m_1=s, m_2=s, m_3=s, m_4=s, m_5=s, \dots}
\quad\mbox{etc}
\nonumber
\ .
\end{eqnarray}
It can be shown that this state is an eigenstate of lowest
energy in one-magnon space ${\mathcal H}(M=Ns-1)$.  On an
extended lattice such as the kagome chain, \figref{F-1-A},
or the two-dimensional kagome lattice, \figref{F-2-B},
many of these objects can be placed in such a way that they do
not interact \cite{SSR:EPJB01,SHS:PRL02}.  Also in this case it
can be shown that these states are eigenstates of lowest energy
in their respective $n$-magnon space.

\section{Consequences}
\label{sec-3}
An immediate consequence of the independence of the localized
one-magnon states and thus of the linear dependence of the
minimal energies $\Emin$ on $M$ is the magnetization jump at the
saturation field $\Bsat$ as well as the high degeneracy of
levels at this field value. In an infinite lattice such as the
kagome or pyrochlore lattice both quantities are macroscopic
\cite{SHS:PRL02,RSH:JPCM03,DeR:PRB04,HoR:CMP05}, although it
turns out that it is rather involved to evaluate the exact
degeneracy at the saturation field due to possible relations
between the $n$-magnon product states \cite{SRM:06}. The
ground-state degeneracy at $\Bsat$ is related to a finite
$(T=0)$ entropy per site, i.e.\ $S(T=0)/N>0$ (for the kagome
chain one obtains $S(T=0)/N \approx 0.1604039$
\cite{DeR:PRB04}).  In the context of magnetocalorics such a
residual entropy gives rise to large adiabatic cooling rates $
\left(\pp{T}{B}\right)_S = -\frac{T}{C}\left(\pp{S}{B}\right)_T
$ in the vicinity of the saturation field
\cite{ZhH:JSM04,DeR:06}. In a real compound, e.g. \mofe, the
perfect degeneracy at $\Bsat$ would be lifted by residual, e.g.
dipolar or Dzyaloshinskii-Moriya interactions. Consequently, the
magnetization jump would be smeared out, compare
Ref.~\cite{SLM:EPL01} for \mofe. Nevertheless, the low-lying
density of states remains large at $\Bsat$, which would still be
clearly visible in magnetocaloric investigations.

\section{Relation to flat bands}
\label{sec-4}
\begin{figure}[h]
\begin{minipage}{18pc}
\begin{center}
\includegraphics[width=14pc]{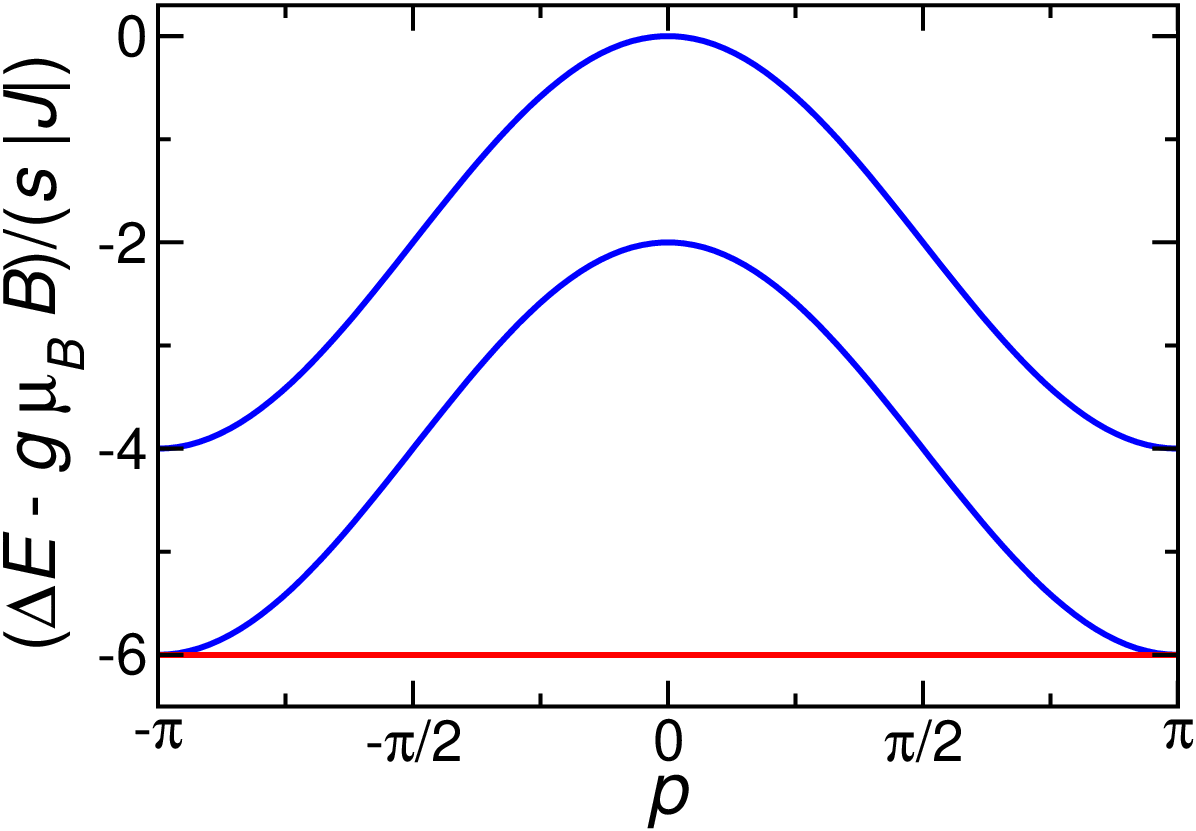}
\caption{\label{F-3-A}One-magnon excitation energies $\Delta E$
  for the kagome chain; the flat band
  consists of $N/3$ degenerate levels.}
\end{center}
\end{minipage} 
\hspace{2pc}%
\begin{minipage}{18pc}
\begin{center}
\includegraphics[width=13pc]{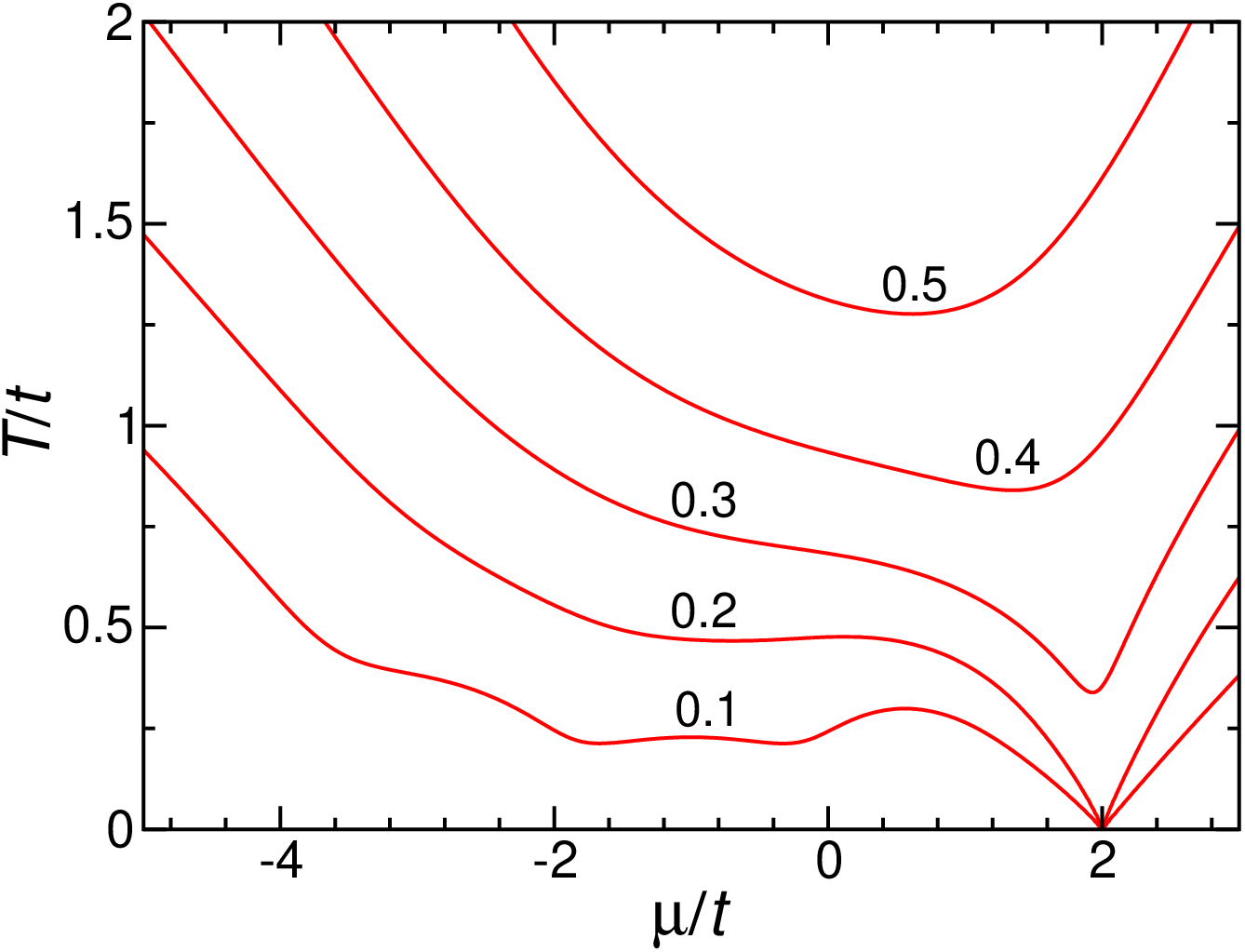}
\caption{\label{F-3-B}Lines of constant
    entropy for free spinless fermions on the kagome chain.
    The value of the entropy per site
    $S(T)/N$ is indicated next to each line.} 
\end{center}
\end{minipage} 
\end{figure}

It is clear that the energetically degenerate independent
localized one-magnon states on translationally symmetric
lattices can be superimposed to form eigenstates of the
translation operator with the same energy which leads to a flat
band. Figure \ref{F-3-A} shows the three energy bands of 
the kagome chain introduced in \figref{F-1-A}.
We observe that one third
of all one-magnon states form independent magnons or equally
well belong to the flat band. Therefore a jump of one third
of the saturation magnetization occurs at $\Bsat$ for
$s=1/2$, see \figref{F-1-B}.

The emergence of flat bands has been already noted in the
context of \emph{line graphs} and \emph{flat-band
  ferromagnetism}, see e.g.
\cite{Mie:JPA92B,Tas:PRL92,MiT:CMP93,Tas:PTP98,IKW:PRB98,NGK:JPSJ03,MKO:JPSJ05}.
A connection can be made by replacing the Heisenberg model by a
Hubbard model \cite{HoR:CMP05}. The roles of the magnetic
exchange $J$ and the applied magnetic field are then played by
the hopping integral $t$ the chemical potential $\mu$,
respectively.  If the flat band is the lowest band, then
non-interacting localized excitations can be constructed for
such fermionic systems in a manner very similar to the
Heisenberg model, and these fermionic systems exhibit similar
thermodynamic properties, but now as function of temperature and
chemical potential. An example is given by \figref{F-3-B} where
the isentropes for free spinless fermions \cite{HoR:CMP05} on
the kagome chain (\figref{F-1-A}) are displayed. At $\mu =
2\,t>0$ the ground state is degenerate with an extensive $(T=0)$
entropy $S(T=0)/N=\ln(2)/3 = 0.231049\ldots$.  The value $\mu =
2\,t$ corresponds to the saturation field $\Bsat$ of the
Heisenberg model, and a very similar behavior would be observed
in its vicinity, e.g. the slopes of the isentropes correspond to
the adiabatic cooling rates \cite{DeR:06}. However, at lower
values of $\mu$ ($B$) there are qualitative differences: free
spinless fermions (whose dispersion is essentially given in
\figref{F-3-A}) have no band gaps, while the magnetization curve
of the $s=1/2$ Heisenberg model presumably has plateaus
(corresponding to gaps) at least at one third and two thirds of
the saturation magnetization (see \figref{F-1-B}).
It may be remarked here that flat-bands on partial line graphs
do not need to be ground-state bands \cite{MKO:JPSJ05}.

\end{document}